\documentclass[aps,apl,a4paper,twocolumn,superscriptaddress,epsf]{revtex4}

\usepackage{graphicx}
\usepackage{bm}
\usepackage{amsmath}
\usepackage{rotating}

\begin{document}

\hyphenation{Schwer-punkt-pro-gramm}

\title{Sensing electric and magnetic fields with Bose-Einstein Condensates}

\author{S.~Wildermuth}
\email[E-Mail: ]{wildermuth@physi.uni-heidelberg.de}
\author{S.~Hofferberth}
\author{I.~Lesanovsky}
\author{S.~Groth}
\affiliation{Physikalisches Institut, Universit\"at Heidelberg,
69120 Heidelberg, Germany}
\author{I.~Bar-Joseph}
\affiliation{Department of Condensed Matter Physics, Weizmann
Institute of Science, Rehovot 76100, Israel}
\author{P.~Kr\"uger}
\author{J.~Schmiedmayer}
\affiliation{Physikalisches Institut, Universit\"at Heidelberg,
69120 Heidelberg, Germany}

\date{\today}

\begin{abstract}
We discuss the application of Bose-Einstein condensates (BECs) as
sensors for magnetic and electric fields. In an experimental
demonstration we have brought one-dimensional BECs close to
micro-fabricated wires on an atom chip and thereby reached a
sensitivity to potential variations of $\sim 10^{-14}$eV at
$3\mu$m spatial resolution. We demonstrate the versatility of this
sensor by measuring a two-dimensional magnetic field map $10\mu$m
above a $100\mu$m-wide wire. We show how the transverse
current-density component inside the wire can be reconstructed
from such maps. The field sensitivity in dependence on the spatial
resolution is discussed and further improvements utilizing
Feshbach-resonances are outlined.
\end{abstract}

\addtolength{\textheight}{+1.5cm}

\maketitle


The measurement of magnetic and electric fields in the immediate
vicinity of surfaces allows to characterize micro-structured
devices. These maps of the local fields provide an insight into
details of the charge distribution and into the transport of the
electron gas in various geometries. This is of great interest for
fundamental studies as well as for technical matters of quality
control of microchips \cite{Cha00}. Electric fields can for
example be probed with high precision by means of single-electron
transistors \cite{Yoo97}. Conventionally available methods for
measuring magnetic fields exist for either high field sensitivity
at low spatial resolution (SQUID \cite{Fal04} and thermal atom
magnetometers \cite{Kom03}) or high resolution at low sensitivity
(MFM \cite{Fre01} and Hall probes \cite{Oral02}).

In this letter we study the performance of a novel field sensor
\cite{Wil05} that is ideally suited for field measurements close
(single microns) to micro-structures. It simultaneously features
high spatial resolution and high field sensitivity. For sensing
the fields, we use one-dimensional Bose-Einstein condensates (BEC)
that we prepare close to the surface of atom chips \cite{Fol02}.
High resolution imaging of these BECs enables us to measure the
density profile of the cold atomic clouds. The variation of both
magnetic and electric fields can be inferred as even slightest
inhomogeneities in these fields measurably alter the trapping
potentials. In demonstration experiments we have reached a
sensitivity to potential variations of $\sim 10^{-14}$eV at a
spatial resolution of $3\mu$m. For magnetic field measurements
this corresponds to a sensitivity of $\sim 10^{-9}$T; we could
detect electric field modulations on the order of V/cm,
corresponding to the field of $\sim 10$ elementary charges at a
distance of $\sim 10\mu$m.

We study in detail how various parameters determine the
sensitivity of the sensor and outline a route to even enhanced
performance. Under ideal conditions, the BECs outperform
conventional devices over a wide spatial resolution range.
Furthermore, we show how a magnetic field map measured near a
conductor can be used to reconstruct the local current profile
inside the conductor.

The basis of our sensor is a trapped highly elongated BEC which
can be precisely positioned microns above a sample to be probed.
We create these BECs in out atom chip apparatus. We start by
collecting ${}^{87}$Rb atoms from a background vapor by an
integrated mirror magneto-optical trap (MOT). The atoms are
transferred to magnetic traps created by micro-fabricated gold
wires mounted on a silicon surface (atom chip) \cite{Fol02,Gro04}.
A quasi 1d BEC ($>1$mm long with aspect ratios up to several
thousand) contains up to $10^5$ atoms in the $F=m_F=2$ state. The
atom number and the desired chemical potential $\mu$ can be
adjusted during the final evaporative cooling stage \cite{Wil04}.

\begin{figure}[b]
    \centering
    \includegraphics[width=\columnwidth]{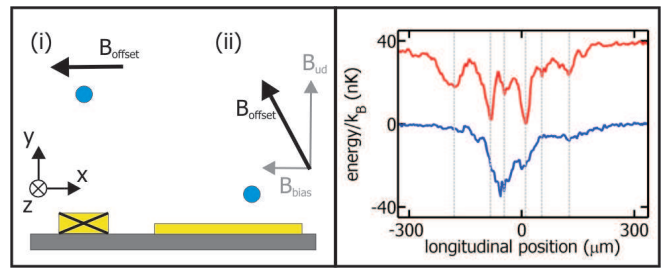}
    \caption{\emph{left}: An elongated BEC held by a trapping wire can be
    arbitrarily positioned above the wire itself (i) or above an
    independent sample to be probed (ii) by freely
    adjusting the current in the wire and the offset-field.
    \emph{right}: Using a BEC held above an independent wire (transverse trap frequency $\omega_{\mathrm{tr}}=2\pi\cdot 700$Hz) the
    potential variations along the z-direction have been measured.
    The blue (lower) curve has been measured with a current passing through the
    wire.
    The red (upper) curve has been measured at the same position but
    charging the wire to amplify the electric potentials.
    The dotted lines are a guide to the eye for comparing the different patterns. The
    red (upper)
    curve has been shifted by $40$nK for visibility.
    }\label{fig1}
\end{figure}

The magnetic micro-trap is created by superimposing the magnetic
field of a current carrying wire with an external homogeneous
offset field $B_{\mathrm{offset}}$ perpendicular to the wire
(Fig.~\ref{fig1}left). This trapping potential allows to
arbitrarily position the minimum of the magnetic trap and thus the
BEC by choosing the current in the wire and the offset field
(magnitude and direction) in an appropriate way. The confinement
in the direction along to the wire (z-direction) is generated by a
varying magnetic field component $B_{\mathrm{z}}$ parallel to the
wire \cite{Fol02}.

The local density of trapped thermal clouds or BECs is imaged in
situ or after ballistic expansion by high resolution ($3\mu$m)
absorption imaging. This density profile $n_{\mathrm{1d}}(z)$ can
be converted into a map of the spatial profile of the longitudinal
potential energy variations $V(z)$. The sensitivity to potential
variations of a thermal atomic cloud is given by their
temperature~$T$: $n_{1d}\sim \exp{(-V/k_BT)}$~\cite{Jon03,Est04}.
In the case of a BEC the relevant energy scale is given by the
chemical potential~\cite{Kra02} which can be orders of magnitude
smaller than the temperature of a thermal cloud ($\mu\ll k_BT$).
We were able to profit from this exceptional potential sensitivity
because atom chips with very low disorder potentials have been
used~\cite{Kru04}.

In elongated traps, the chemical potential $\mu$ of the BEC is of
the order of the energy level spacing in the transverse (strong
confinement) direction ($\mu\simeq\hbar\omega_{\mathrm{tr}}$), but
still much larger than the energy level spacing in the
longitudinal (weak confinement) direction
($\mu\gg\hbar\omega_{\mathrm{lo}}$). In this cross-over region the
three-dimensional regime is smoothly connected to the
one-dimensional mean-field regime by a local density approximation
\cite{Ger04}. The longitudinal potential can be calculated from
the 1d-density profile by
\begin{equation}
V_0+V(z)=-\hbar\omega_{\mathrm{tr}}\sqrt{1+4a_{\mathrm{scat}}n_{\mathrm{1d}}(z)}\label{pot}
\end{equation}
where $a_{\mathrm{scat}}$ is the s-wave scattering length ($5.2$nm
for ${}^{87}$Rb \cite{Kem02}) and $V_0$ is an arbitrary offset.
The sensitivity to potential variations is proportional to
$\hbar\omega_{\mathrm{tr}}$: Weaker confinement allows to see
smaller potential variations.

The optimal potential single shot sensitivity $\Delta V$ of a BEC
as field sensor is reached by detecting the density distribution
atom shot-noise limited. The desired spatial resolution is not
necessarily equal in the longitudinal ($z_0$) and transverse
($\rho_0$) directions. Ideally, the trap parameters are chosen
such that the transverse ground state size matches $\rho_0$. In
this optimal situation the sensitivity of a one-dimensional BEC in
the mean-field regime is given by:
\begin{equation}
\Delta V=\frac{\gamma \Delta N}{\rho_0^2 z_0}, \quad
\gamma=\frac{2\hbar^2}{m}a_{\mathrm{scat}}\label{sens}
\end{equation}
For our detection imaging noise of $\Delta N\sim
4\mathrm{atoms}/\mathrm{pixel}$ a single-shot single-point
sensitivity to potential variations of $\sim 10^{-13}$eV ($\sim
10^{-14}$eV) can be reached at $\omega_{\mathrm{tr}}=2\pi\cdot
3$kHz ($\omega_{\mathrm{tr}}=2\pi\cdot 300$Hz). Variations in the
longitudinal potential can originate from local magnetic as well
as from electric fields (Fig.~\ref{fig1} right).

The sensitivity to electric field modulations can be enhanced by
adding a homogeneous electric offset-field $E_0$~\cite{Mcg04}. In
this case the potential is related to the electric field by
$V(z)\approx -\alpha E_0E(z)$ where $\alpha$ is the
polarizability. For our atom detection a field sensitivity of
$\Delta E=0.4$V/cm can be reached at
$\omega_{\mathrm{tr}}=2\pi\cdot 3$kHz and $E_0=1$kV/cm. This
equals the field strength produced by a fraction of an elementary
charge ($\sim 25\%$) detected at a distance of $3\mu$m.

\begin{figure}
    \centering
    \includegraphics[width=\columnwidth]{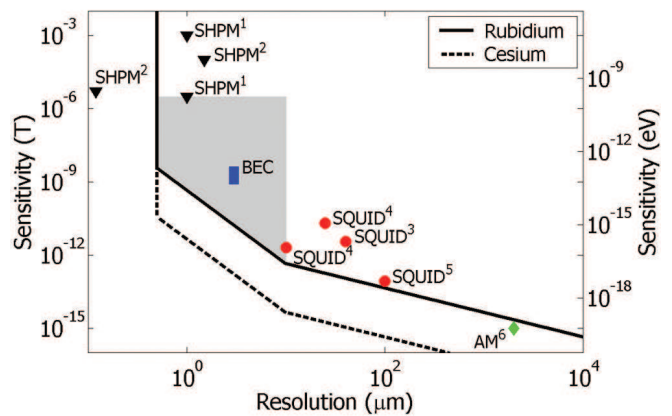}
    \caption{The potential sensitivity versus spatial resolution
    for the BEC-sensor has been plotted according to
    Eq.~(\ref{sens}). The solid curve shows the sensitivity of the demonstrated
    sensor using a Rubidium BEC. Tuning the scattering length of
    the atoms by means of a Feshbach resonance leads to even
    higher sensitivity, e.g. the estimated sensitivity of a Cesium BEC
    has been plotted as dashed curve. In comparison the
    field sensitivity versus spatial resolution for
    state-of-the-art magnetic microscopes is shown (Scanning Hall
    Probe Microscopy ${}^1$\cite{Bro03}, ${}^2$\cite{Oral02},
    Superconducting Quantum Interference Device ${}^3$\cite{Bau03}, ${}^4$\cite{Fal04},
    ${}^5$\cite{Zha02}, and thermal atom magnetometer ${}^6$\cite{Kom03}).
    The dark grey shaded region indicates the sensitivity-resolution range
    currently accessible only to the demonstrated BEC sensor.}
    \label{fig2}
\end{figure}

In the case of local magnetic field variations~\cite{Kru04} the
longitudinal potential is related to the magnetic field by
$V(z)=m_Fg_F\mu_B B(z)$ where $\mu_B$ is Bohr's magneton, $m_F$ is
the quantum number associated with the Zeeman state of the atom
and $g_F$ is the Lande-factor. To compare the performance of the
BEC-sensor to commonly used magnetic field detectors
(Fig.~\ref{fig2}) the dependence of the field sensitivity on the
spatial resolution of the measurement has to be taken into
account.

Two regimes can be distinguished which are depicted in
Figure~\ref{fig2}: The field sensitivity scales most strongly with
the spatial resolution of the measurement if $z_0=\rho_0$. This
can be achieved if the transverse ground-state size matches the
desired resolution and if the resolution of the imaging system is
better than $z_0$. By using a transition in the blue (around
$\lambda=421.67$nm for the Rubidium
$5^2\mbox{S}_{1/2}-6^2\mbox{P}_{1/2}$ resonance~\cite{Ito00}) an
imaging resolution of $\sim 500$nm can be achieved. Consequently,
the spatial resolution of the magnetic field measurement in this
range ($\Delta s_B=0.5-10\mu$m) can be assumed to be limited by
the imaging system only. Since generation of BECs in traps which
are shallower than $\omega =2\pi\cdot 1$Hz (more than $10\mu$m
ground-state size) seams to be challenging~\cite{Lea03},
sensitivity can further be enhanced by decreasing $z_0$ at a fixed
transverse confinement of $2\pi\cdot 1$Hz. This leads to an
anisotropic spatial resolution of the sensor.

The sensitivity is given by the chemical potential for a specific
longitudinal 1d-density and its shot-noise. Decreasing the
strength of the interactions between the atoms leads to a lower
chemical potential at the same $n_{1d}$ and consequently higher
sensitivity (Eq.~\ref{sens}). The sensor can be operated at an
arbitrary homogeneous field $B_z$; the s-wave scattering length
$a_{\mathrm{scat}}$ can be adjusted using Feshbach resonances. In
the case of Cesium, $a_{\mathrm{scat}}$ becomes zero around an
easily manageable magnetic field value of $17.0$G and the
linearized slope around this zero-crossing of $a_{\mathrm{scat}}$
can be estimated to be $\sim 6$nm/G~\cite{Vul99}. Additionally the
higher mass of Cesium compared to Rubidium further increases the
sensitivity. Decreasing the scattering length to $0.1$nm which
requires an easily manageable control of the magnetic field of
$15$mG would gain a factor of $\sim 100$ in sensitivity (dashed
line in Fig.~\ref{fig2}).

\begin{figure}
    \centering
    \includegraphics[width=\columnwidth]{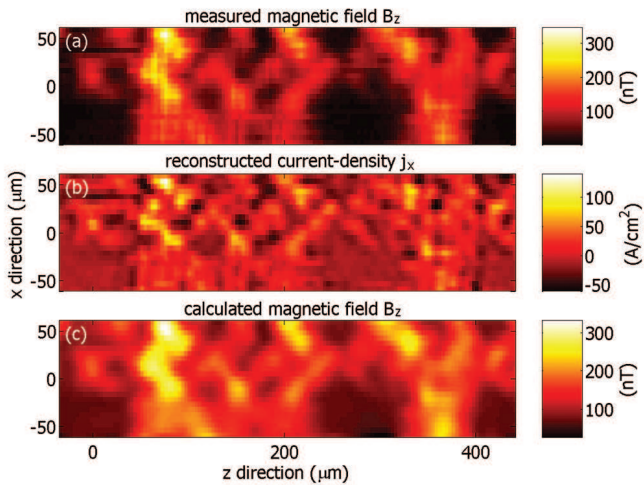}
    \caption{(a) A two-dimensional scan of the z-component of the
    magnetic field has been taken $10\mu$m above a current-carrying
    gold wire (cross-section $100\times 3.1\mu\mbox{m}^2$) at a
    homogeneous offset-field of $20$G and a current of $340$mA
    resulting in a transverse trap frequency of
    $\omega_{\mathrm{tr}}=2\pi\cdot 3$kHz. This field map has been
    obtained by positioning the BEC transversely at $28$ equally
    spaced positions. (b) The underlying current density has been
    reconstructed according to Eq.~(\ref{cur}). This map clearly shows
    that not only current flow deviations caused by wire-edge
    roughness are important. Local properties of the bulk are more
    dominant. (c) Inserting the reconstructed current density back
    into Biot-Savart's law yields the initial varying field. The
    visible smoothing arises from filtering the experimental data in
    Fourier space.}\label{fig3}
\end{figure}

As an application of our potential imaging we have measured a map
of the local longitudinal magnetic field variation $10\mu$m above
a flat (cross-section $100\times 3.1\mu\mbox{m}^2$)
current-carrying wire (Fig.~\ref{fig3}a) consisting of 28 equally
spaced positions along the transverse direction of the wire. From
this map we have reconstructed the local transverse current flow
in the conductor.

This reconstruction involves the deconvolution of Biot Savart's
law which can be performed if the current density is assumed to be
confined to a 2d-plane and if boundary conditions on the geometry
of the wire are assumed. In this case only $j_x$ contributes to
the magnetic field $B_{\mathrm{z}}$ and can be calculated
according to
\begin{equation}
j_x(x,z)={\cal F}^{-1}\{
\bar{B}(k_x,k_z)e^{k|y|}\}(x,z)\label{cur}
\end{equation}
where $\bar{B}(k_x,k_z)= {\cal F}\{ B_{z}(x,z)\}(k_x,k_z)$ and
${\cal F}$ indicates a two-dimensional Fourier transform.

In the example presented here two effects have to be taken into
account when estimating the spatial resolution $\Delta s_j$ of the
reconstructed current-density: (i) The magnetic field has been
measured at a finite distance of $y=10\mu$m above the surface
resulting in a smoothing of the magnetic field variations. (ii)
The field has been mapped on a grid given by the spatial
resolution of the imaging system in the z-direction and by the
transverse positioning of the BEC in the x-direction. Here, two
regimes have to be considered: If the spatial resolution $\Delta
s_B$ of the magnetic field measurement is coarser than the
distance $y$ to the wire, one finds $\Delta s_j\approx \Delta
s_B$. In the opposite limit ($y\ge\Delta s_B$) the resolution of
the current-density can be found as follows: Two point-like
current-density components positioned at a distance $\Delta s_j$
can be linked by Biot-Savart's law to the length scale $\Delta
s_B$ of the modulation of the resulting magnetic field, yielding
$\Delta s_j\approx y+\frac{3}{10}\frac{\Delta s_B^2}{y}$. In our
measurement of the $100\mu$m-wide wire, the spatial resolution of
$j_x$ is limited by the distance to the wire. To reveal more
details the atom-surface distance in the field measurement has to
be reduced. It has been shown that a surface approach of single
microns is possible~\cite{Kru04}. Constraints are the limited trap
depth in the presence of attractive surface
potentials~\cite{Lin04} and the finite trap lifetime~\cite{Hen01}.
Trap lifetimes have to be sufficiently long for the condensate to
reach thermodynamic equilibrium. This is typically not problematic
as lifetimes on the order of seconds can be achieved down to
single microns from insulators~\cite{Sin05} or thin conducting
layers ~\cite{Zha05} that are prone to be probed by BEC sensors.

If the current density is reconstructed using Eq.~(\ref{cur}),
high frequency noise in the magnetic field map resulting from the
imaging process has to be taken into account. This noise causes
artificial structures which do not represent the actual current
density distribution. Using the estimation of the expected current
density resolution discussed above, a filter function can be
designed: $F^{-1}(k_x,k_z)=\left[
1+\exp\left(\frac{k_x-c_x}{s_x}\right) \right]\left[
1+\exp\left(\frac{k_z-c_z}{s_z}\right) \right]$. This has been
applied to $\bar{B}(k_x,k_z)$ before computing $j_x$
(Fig.~\ref{fig3}b) using Eq.~(\ref{cur}). A reasonable choice of
the filter parameters is $c_x=0.32\mu\mbox{m}^{-1}$,
$c_z=0.22\mu\mbox{m}^{-1}$, and $s_x=s_z=0.05\mu\mbox{m}^{-1}$. As
a test of the applied deconvolution methodology, $B_z$ has been
calculated from the reconstructed $j_x$ using Biot-Savart's law
directly (Fig.~\ref{fig3}c). Good agreement between this
calculated magnetic field and the measured magnetic field map is
found.

In conclusion, we have shown that high resolution potential images
can be derived from quasi 1d BECs used as sensor.  Imaging the
condensate 1d density one obtains a high resolution ($\sim\mu$m)
potential map along a line on a mm scale. Our experiments
demonstrate the unique sensitivity of the 1d BEC when applied to
magnetic fields, surpassing conventional measurement methods by
orders of magnitude for spatial resolutions in the $\mu$m-range.
As an application of this field sensor we have demonstrated that
BECs can be used to reconstruct the current-density in a
micro-fabricated wire. In future experiments a BEC sensor could be
used to obtain a deeper understanding of the local current flow
for example in superconductors and two-dimensional electron gases.

This work was supported by the European Union, contract numbers
IST-2001-38863 (ACQP), HPRN-CT-2002-00304 (FASTNet),
HPMF-CT-2002-02022, and HPRI-CT-1999-00114 (LSF) and the Deutsche
Forschungsgemeinschaft, contract number SCHM 1599/1-1.

\end{document}